\input psfig.sty
\documentstyle[aps,prl,preprint,floats,epsf]{revtex}

\begin{document}

\preprint{\tighten\vbox{\hbox{\hfil CLNS 97/1460}
                        \hbox{\hfil CLEO 97-1}
}}

\title{Studies of the Cabbibo-suppressed decays 
$D^+\rightarrow\pi^0\ell^+\nu$ and $D^+\rightarrow\eta e^+\nu_e$}

\author{CLEO Collaboration}
\date{\today}

\maketitle
\tighten

\begin{abstract}
Using 4.8 fb$^{-1}$ of data taken with the CLEO II detector, 
the branching fraction for the Cabibbo 
suppressed decay $D^+\rightarrow\pi^0\ell^+\nu$ measured relative to the 
Cabibbo favored decay $D^+\rightarrow\overline{K^0}\ell^+\nu$ 
is found to be 
$0.046\pm 0.014\pm 0.017$.  Using $V_{cs}$ and $V_{cd}$ from 
unitarity constraints, we determine 
$\vert f_+^{\pi}(0)/f_+^K(0)\vert^2=0.9\pm 0.3\pm 0.3$  
We also present a 90\% confidence level upper 
limit for the branching ratio of the decay 
$D^+\rightarrow\eta e^+\nu_e$ relative 
to that for $D^+\rightarrow\pi^0 e^+\nu_e$ of 1.5.  
\end{abstract}
\newpage
{
\renewcommand{\thefootnote}{\fnsymbol{footnote}}

\begin{center}
J.~Bartelt,$^{1}$ S.~E.~Csorna,$^{1}$ V.~Jain,$^{1}$
S.~Marka,$^{1}$
A.~Freyberger,$^{2}$ R.~Godang,$^{2}$ K.~Kinoshita,$^{2}$
I.~C.~Lai,$^{2}$ P.~Pomianowski,$^{2}$ S.~Schrenk,$^{2}$
% ...
\end{center}
 
\small
\begin{center}
$^{1}${Vanderbilt University, Nashville, Tennessee 37235}\\
$^{2}${Virginia Polytechnic Institute and State University,
Blacksburg, Virginia 24061}\\
% ...
\end{center}
 
\setcounter{footnote}{0}
}
\newpage

Interpretation of semileptonic decays of charm mesons is theoretically 
straightforward.  Amplitudes of decay modes are 
proportional to the CKM matrix elements and the form factors describing 
the strong interactions between the final state quarks.  
In this paper we study the Cabibbo suppressed decays 
$D^+\rightarrow\pi^0\ell^+\nu$ 
and $D^+\rightarrow\eta\ell^+\nu$ by measuring the ratios 
$R_{\pi}=B(D^+\rightarrow\pi^0\ell^+\nu)/
B(D^+\rightarrow\overline{K^0}\ell^+\nu)$ and 
$R_{\eta}=B(D^+\rightarrow\eta\ell^+\nu)/
B(D^+\rightarrow\pi^0\ell^+\nu)$.  
Throughout this paper charge conjugate states 
are implied. 

The ratio $R_{\pi}$ is proportional to the product 
$\vert f_+^{\pi}(0)/f_+^K(0)\vert^2
\vert V_{cd}/V_{cs}\vert^2$ where $f_+^P(q^2)$ is the hadronic form factor 
for the decay into the pseudoscalar $P$.  
Since unitarity constraints on the Cabibbo-Kobayashi-Maskawa (CKM) 
matrix\cite{ref1} yield $V_{cd}$ and $V_{cs}$ with good precision, 
measurements of $R_{\pi}$ and $R_{\eta}$ 
mainly provide constraints on the form factor ratios 
$\vert f_+^{\pi}(0)/f_+^K(0)\vert$ and 
$\vert f_+^{\eta}(0)/f_+^{\pi}(0)\vert$.  
Model predictions\cite{ref2} for the first form factor ratio 
range from 0.7 to 1.4.  This range demonstrates the difficulty in predicting 
how the $D$ meson will couple to the light quark combinations.   
Assuming a monopole 
form\cite{ref3} for the $q^2$ dependence of the form factor, the 
decay rate for the decay $D^+\rightarrow P\ell^+\nu$ can be 
written as  
\begin{equation}
\Gamma = c_1^2 |V_{cd}|^2|f^P_+(0)|^2\int
\frac{p^3_P}{(1-\frac{q^2} {M^*})^2}dq^2,
\end{equation}
where $q^2$ is the hadronic four momentum transfer.  The mass of the 
nearest vector pole is $M^*=M_{D^*}$ for $D^+\rightarrow\pi^0\ell\nu$ and 
$D^+\rightarrow\eta\ell\nu$, and $M^*=M_{D_s^*}$ for 
$D^+\rightarrow\overline{K^0}\ell\nu$.  The factor $c_1^2$ accounts for the 
$d\overline{q}$ content of the final state meson P, and is 1/2 for the 
$\pi^0$ and $\eta$ modes ($d\overline{d}$), and 1 for the $\overline{K^0}$ 
mode ($d\overline{s}$).  
There are several models that predict these rates\cite{ref4,ref5}.
Using the framework of Heavy Quark Effective Theory and symmetry arguments,  
measured form factors from semileptonic charm decays can be compared to those 
for the appropriate $b\rightarrow u$ decays\cite{ref6} used to extract 
$\vert V_{ub}/V_{cb}\vert$.  

While the Cabibbo-favored modes in charm semileptonic decay have been 
well measured\cite{ref1,ref7}, there are relatively few measurements 
of Cabibbo-suppressed semileptonic decays.  Previous CLEO results for 
the ratio $R_{\pi}$\cite{ref8} are based on a  
total luminosity of 2.1 fb$^{-1}$, and are superceded by the 
results presented in this paper.   The ratio of branching 
fractions $R_-=B(D^0\rightarrow\pi^-\ell^+\nu)/
B(D^0\rightarrow K^-\ell^+\nu)$ is related 
to $R_{\pi}$ by isospin ($R_{\pi}=0.5R_-$).  Mark III\cite{ref9}, 
Fermilab E687\cite{ref10},  and CLEO\cite{ref11} 
have reported results for $B(D^0\rightarrow\pi^-\ell^+\nu)$ giving a current 
world average for $R_-=0.102^{+0.017}_{-0.016}$.

The data sample used for this analysis was recorded with the CLEO-II 
detector\cite{ref12} operating at the CESR storage ring at 
Cornell University.  A total 
luminosity of 4.8 $\mathrm{fb^{-1}}$ of 
$e^+e^-$ collisions was recorded at the 
$\Upsilon(4S)$ resonance and in the continuum nearby.  

In $D^+$ decays, the combinatoric background can be suppressed by 
requiring that the $D^+$ be  
produced in the decay chain $D^{*+}\rightarrow D^+\pi^0$.
The CLEO-II detector, with its 
excellent photon detection efficiency, is ideally suited for detecting 
the neutral pions from this decay.  Because the final state neutrino is 
not detected in semileptonic decays, we define 
$\delta m= M_{\pi^0_Sh_F\ell^+}-M_{h_F\ell^+}$, where $h_F$ refers to the 
$D^+$ daughter meson, the ``fast'' $\pi^0$ ($\pi^0_F$), 
the $\overline{K^0}$, or the $\eta$.  The $\pi^0_S$ refers to the 
``slow'' $\pi^0$ from the $D^{*+}$, which is constrained by the 
production and decay kinematics to have a momentum less than 0.4 GeV/$c$.  
While the peak in $\delta m$ is not as narrow as the peak in fully 
reconstructed hadronic $D^+$ decays, a definite peak remains.  The width 
of the peak in this distribution increases as more energy is 
carried by the neutrino.  We therefore limit the neutrino energy by 
requiring $1.4\le M_{h_F\ell^+}<1.8$ GeV/c$^2$.  

Electrons with momenta above 0.7 GeV/c are identified by requiring that 
the ratio of the energy (E) deposited in the CsI calorimeter and the 
momentum (p) measured in the tracking system, E/p, be close to unity and that 
the energy loss measured by the tracking system be consistent with the 
electron hypothesis.  Muons with momenta above 1.4 GeV/c are identified by 
their ability to penetrate five nuclear interaction 
lengths.  Electrons (muons) within the fiducial volume are 
identified with an efficiency of 94\% (93\%).  The probability of a hadron 
being misidentified as a lepton is $(0.20\pm 0.06)\%$ for electrons and 
$(1.4\pm 0.2)\%$ for muons.  We require the leptons 
to be found in the central region of the detector, where the 
resolution is best and the acceptance well-understood.

Isolated photons detected by the CsI calorimeter with a minimum energy of 
30 MeV are paired to form $\pi^0$ and $\eta$ candidates.  For the slow pion, 
the $\gamma\gamma$ mass is constrained to be within 
$2.5$ standard deviations (about 12.5 MeV/c$^2$) of the nominal 
$\pi^0$ mass.   
For the fast $\pi^0$ ($\eta$), the reconstructed mass is required to 
be within the range 0.105-0.165 GeV/c$^2$ (0.510-0.585 GeV/c$^2$).  
The decay channel 
$\eta\rightarrow\pi^+\pi^-\pi^0$ 
was not considered because of its low reconstruction efficiency.  
For the normalizing $D^+\rightarrow\overline{K^0}\ell^+\nu$ mode, 
we identify the $\overline{K^0}$ through the 
$\pi^+\pi^-$ decay of its $K_S$ 
component. We require the $\pi^+\pi^-$ pair to form a secondary vertex of 
the correct mass that is displaced at least four standard deviations 
from the primary vertex.   

Combinatoric backgrounds are reduced by several means.  We impose the 
kinematic criteria $0.175\le p_{\pi^0_S}<0.350$ GeV/$c$, 
$p_{h_F}\ge 0.7$ GeV/$c$, and 
$\vert\vec{p}_{h_F}+\vec{p}_{\ell}\vert\ge 2.1$ GeV/$c$.  Backgrounds from 
B meson decay are reduced by requiring that the ratio of Fox-Wolfram 
moments\cite{ref13} $R_2= H_2/H_0$ satisfy $R_2\ge 0.2$.  Finally, we 
consider only well-measured tracks and events with a hadronic event 
structure.

Backgrounds can be divided into four classes: fake slow pions (fake $D^*$s), 
fake fast hadrons, fake leptons, and uncorrelated 
fast-hadron, lepton pairs (fake $D^+$s).  The major contribution to 
the fake $D^+$ background in the $D^+\rightarrow\pi^0\ell^+\nu$ channel 
comes from feeddown from $D^+\rightarrow\overline{K^0}\ell^+\nu$, 
$\overline{K^0}\rightarrow\pi^0\pi^0$.  We can correct for this 
background knowing only the ratio of the reconstruction efficiency 
for $D^+\rightarrow\pi^0\ell^+\nu$ to the efficiency to reconstruct 
$D^+\rightarrow\overline{K^0}\ell^+\nu$, 
$\overline{K^0}\rightarrow\pi^0\pi^0$ as $\pi^0\ell^+\nu$, 
which we determine 
from Monte Carlo simulation.  Monte Carlo studies indicated that the 
feedthrough from other semileptonic charm decays and from $B\overline{B}$ 
events is negligible.  The other background components were determined from 
fits to the data. 

\begin{figure}
\centerline{\psfig{figure=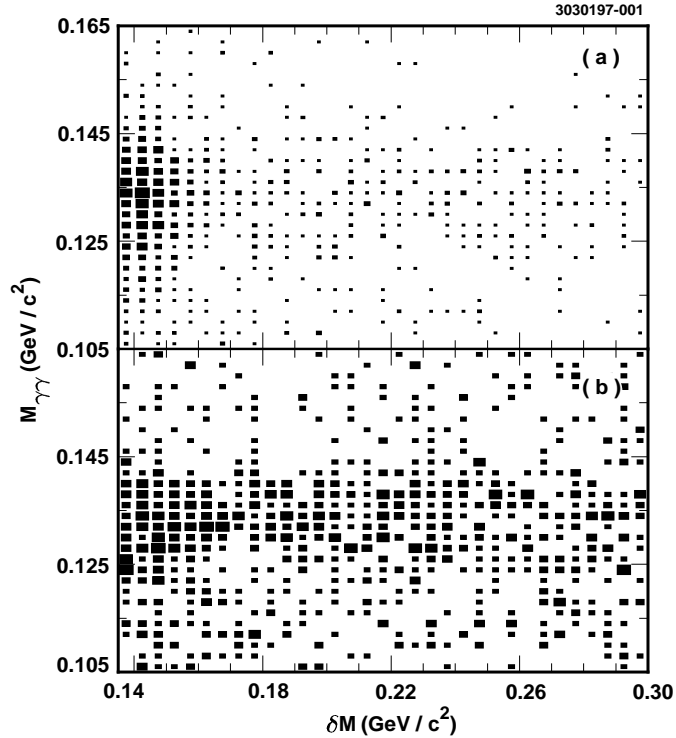,width=3.5in}}
%\vspace{8.5 cm}
%\special{psfile=/data1/bean/pi0/p0f1.ps hscale=60 vscale=35 
%hoffset=20 voffset=-10}
\caption{The distribution of $M_{\gamma\gamma}$ versus 
$\delta m$ for a) $D^+\rightarrow\pi^0e^+\nu$ Monte Carlo events and 
b) data.}
\label{fig1}
\end{figure}

\begin{figure}
\centerline{\psfig{figure=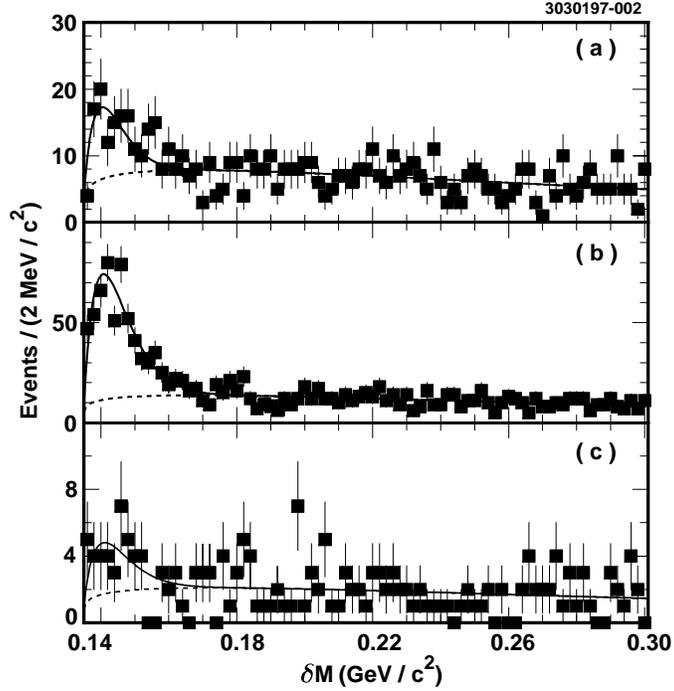,width=3.5in}}
%\vspace{8.5 cm}
%\special{psfile=/data1/bean/pi0/p0f2.ps hscale=60 vscale=35 
%hoffset=20 voffset=-10}
\caption{The $\delta m$ spectra for 
data passed through the a) $D^+\rightarrow\pi^0e^+\nu_e$ analysis with 
$0.115\leq M_{\pi^0_F}<0.153$ GeV/c$^2$, 
b) $D^+\rightarrow\overline{K^0}e^+\nu_e$ analysis with 
$0.48\leq M_{K^0_S}<0.52$ GeV/c$^2$, and c) 
$D^+\rightarrow\eta e^+\nu_e$ analysis 
with $0.51\leq M_{\eta}<0.58$ GeV/c$^2$.  
The solid line indicates the total fit while the 
dashed line indicates the background function.}
\label{fig2}
\end{figure}

\begin{figure}
\centerline{\psfig{figure=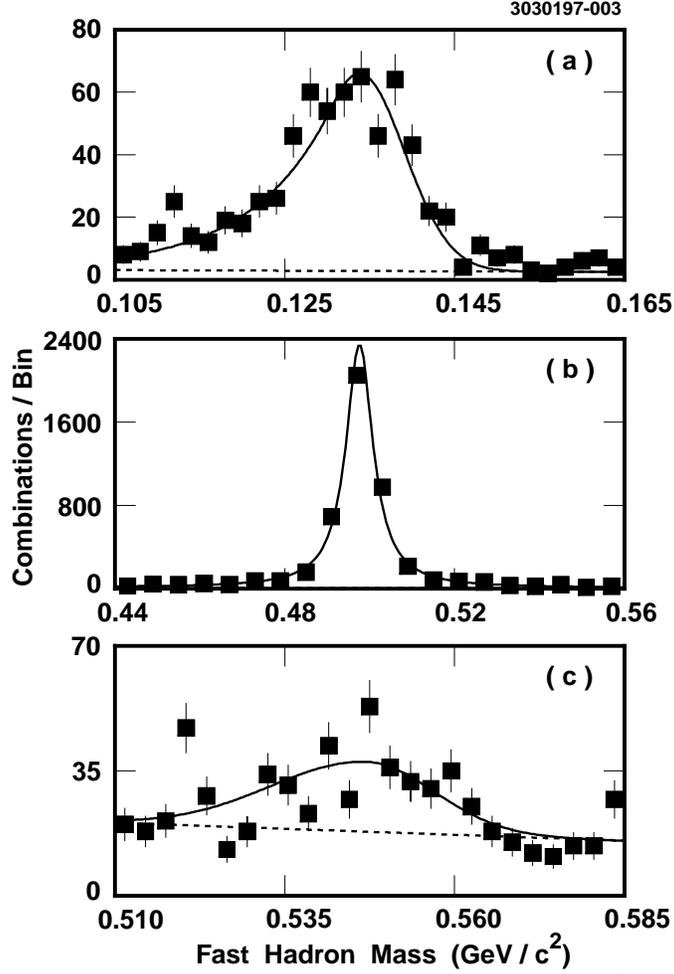,width=3.5in}}
%\vspace{8.5 cm}
%\special{psfile=/data1/bean/pi0/p0f3.ps hscale=60 vscale=35 
%hoffset=20 voffset=-10}
\caption{The a) $M_{\gamma\gamma}=M_{\pi^0_F}$, b) $M_{\pi^+\pi^-}$, and 
c) $M_{\gamma\gamma}=M_{\eta}$ spectra for 
data passed through the $D^+\rightarrow\pi^0e^+\nu_e$, 
$D^+\rightarrow\overline{K^0}e^+\nu_e$, and 
$D^+\rightarrow\eta e^+\nu_e$ analyses, respectively, 
with $\delta m<0.3$ GeV/c$^2$. 
 The solid line 
indicates the total fit while the dashed line indicates the 
background function.}
\label{fig3}
\end{figure}

We fitted the two dimensional  
distribution of $\delta m$ versus fast hadron 
mass to extract the signal yield.  Figure\ \ref{fig1} shows 
the distributions for the signal Monte Carlo and data for the 
$D^+\rightarrow\pi^0e^+\nu_e$ mode.  Figure\ \ref{fig2} shows the 
$\delta m$ projection for the $\pi^0$, $\overline{K^0}$, and $\eta$ modes.  
Figure\ \ref{fig3} shows the fast hadron mass distributions.  The 
fits used a parametrization of the fast hadron mass obtained by fitting 
these one-dimensional projections.  The signal shape in $\delta m$ was 
determined from fits to the distributions of reconstructed signal Monte 
Carlo.  The fake lepton background was determined by performing a fit to the 
distributions of events which satisfied all requirements except for the 
lepton identification requirement.  The signal yields from these fits were 
then scaled by the measured misidentification probabilities and subtracted 
from the yields from the fit to the data.  The parameterization of the 
fake $D^*$ background in $\delta m$ was determined by looking at a sample of 
data events whose fast hadron mass was more than 4 sigma from the nominal 
mass.  The signal yields, fake lepton 
yields, and signal reconstruction efficiencies are presented in 
Table\ \ref{tab1}.  The efficiencies were determined from fits to the 
distributions from samples of reconstructed signal Monte Carlo.

With the results from the fits given in Table\ \ref{tab1}, 
we proceed to calculate the ratio of branching fractions 
$R_{\pi}=[B(D^+\rightarrow\pi^0\ell^+\nu)]/
[B(D^+\rightarrow\overline{K^0}\ell^+\nu)]$.  
For each leptonic mode we define,  
$$R_{\pi}=\frac{N(\pi^0_F\ell^+\nu)}{N(K^0_S\ell^+\nu)}
\frac{\epsilon_{\overline{K^0}\ell^+\nu}(\overline{K^0}\ell^+\nu)}
{\epsilon(\pi^0\ell^+\nu)} - 
\frac{\epsilon_{\overline{K^0}\ell^+\nu}(\pi^0\ell^+\nu)}
{\epsilon(\pi^0\ell^+\nu)}$$
Here $N(\pi^0_F\ell^+\nu)$ and $N(K^0_S\ell^+\nu)$ are the 
two signal yields after 
background subtraction, $\epsilon(\pi^0\ell^+\nu)$ is the efficiency for 
a $\pi^0\ell^+\nu$ decay to be reconstructed as itself, 
$\epsilon_{\overline{K^0}\ell^+\nu}(\overline{K^0}\ell^+\nu)$ 
is the efficiency for 
a $\overline{K^0}\ell^+\nu$ decay to be reconstructed as 
$\overline{K^0}\ell^+\nu$, 
and $\epsilon_{\overline{K^0}\ell^+\nu}(\pi^0\ell^+\nu)$ is the efficiency for 
a $\overline{K^0}\ell^+\nu$ decay to be reconstructed as 
$\pi^0\ell^+\nu$.  
The ratio for 
electrons was found to be 
$R_{\pi}=(4.5\pm 1.6\pm 1.9)\%$, where the first error is 
statistical and the second is systematic.  The ratio for   
muons was found to be $R_{\pi}=(4.8\pm 3.1\pm 3.2)\%$.  
Here the error from fake muon subtraction is 
substantial and the detection efficiency is lower than for the electron 
channel.  
We combine the results weighted by their errors to find 
$R_{\pi}=(4.6\pm 1.4\pm 1.7)\%$.  

Most of the systematic effects 
cancel in the ratio of branching fractions because we impose similar 
requirements on both the signal and normalization modes.  
The systematic error for the electron channel 
is dominated by the parameterizations of 
the shapes in the $\delta m$ distribution (30\%).  This error is correlated 
between the $\pi^0e^+\nu_e$ and $\overline{K^0}e^+\nu_e$ channels.      
The systematic error in the ratio due to Monte Carlo 
simulations of $K^0_S\rightarrow\pi^+\pi^-$ and 
$\pi^0_F\rightarrow\gamma\gamma$ is 
conservatively placed at 10\%.  Other systematic errors for the 
electron channel include: 
statistical error on efficiency fits from Monte Carlo samples (7\%), 
fake lepton subtraction (7\%), $D^+\rightarrow\overline{K^0}e^+\nu_e$ feeddown 
(9\%), other semileptonic charm decay feeddown 
(16\%), and $B\overline{B}$ feeddown (13\%).  The systematic errors are 
added in quadrature to obtain a total systematic error in the ratio for 
electrons of 41\%.

The fit to the $D^+\rightarrow\eta e^+\nu$ channel yielded $6\pm 8$ 
events.  We did not consider the muon channel due to the low 
detection efficiency.  To obtain an upper limit on $R_{\eta}$, we 
scale this yield by the reconstruction efficiency of  
$(0.26\pm 0.02)\%$, and normalize to the average 
$D^+\rightarrow\pi^0\ell^+\nu$ yield of $(4.39\pm 2.22)\times 10^3$ events.
The latter was estimated from our $R_\pi$ measurement and the  
average of the efficiency-corrected yields for 
$D^+\rightarrow\overline{K^0}\ell^+\nu$ in the electron and muon channels.  
We find $R_{\eta}={B(D^+\rightarrow\eta\ell^+\nu)\over 
B(D^+\rightarrow\pi^0\ell^+\nu)} =
 < 1.5.$ at the 90\% confidence level.  
This result is dominated by statistical error, but includes a 
30\% systematic error that was combined in quadrature with the 
statistical error. 

We have measured the branching fraction of the Cabibbo suppressed decay 
$D^+\rightarrow\pi^0\ell^+\nu$ relative to 
$D^+\rightarrow\overline{K^0}\ell^+\nu$.  Using 
our measurement of this ratio, we find using Equation (1)    
$\vert f_+^{\pi}(0)/f_+^K(0)\vert^2\vert V_{cd}/V_{cs}\vert^2 = 
0.046\pm 0.014\pm 0.017$.  The integral in Equation (1) times the constant 
term is approximately 1 here.   
Unitarity constraints on the CKM matrix yield 
$\vert V_{cd}/V_{cs}\vert^2=0.051\pm 0.001$\cite{ref1} which translates to 
a value of $0.9\pm 0.3\pm 0.3$ for $\vert f_+^{\pi}(0)/f_+^K(0)\vert^2$.  
Model predictions\cite{ref2} are in 
agreement with our measurement.  We can combine our measurement of  
$R_{\pi}$ with the measurements of 
$0.5\times R_-$ to obtain $R_{\pi}=0.050\pm 0.008$ and 
$\vert f_+^{\pi}(0)/f_+^K(0)\vert=0.99\pm 0.08$.  
The upper limit on the ratio $R_{\eta}$ 
is consistent with current predictions.

We gratefully acknowledge the effort of the CESR staff in providing us with
excellent luminosity and running conditions.
J.P.A., J.R.P., and I.P.J.S. thank                                           
the NYI program of the NSF, 
M.S. thanks the PFF program of the NSF,
G.E. thanks the Heisenberg Foundation, 
%I.P.J.S. and T.S. thank the 
%
% TNLRC for Ian and Tomasz will expire Jan. 1 1995
%
%TNRLC, 
K.K.G., M.S., H.N.N., T.S., and H.Y. thank the
OJI program of DOE, 
J.R.P., K.H., M.S. and V.S. thank the A.P. Sloan Foundation,
and A.W. and R.W. thank the 
Alexander von Humboldt Stiftung
for support.
%
% add Cottrell Scholar of Research Corporation for M.S. - Sept. 96
M.S. is supported as a Cottrell Scholar of Research Corporation.
This work was supported by the National Science Foundation, the
U.S. Department of Energy, and the Natural Sciences and Engineering Research 
Council of Canada.

\begin{table}
\caption{Results of fits to the $M_{h_F}$ versus $\delta m$ 
distributions for each of the three analyses.}
\label{tab1}
{\small 
\begin{tabular}{ccccc}
Sample & $\pi^0\ell^+\nu$ & & $\overline{K^0}\ell^+\nu$ & \\ 
 & Electrons & Muons & Electrons & Muons \\ \hline
Data & $75\pm 15$ & $83\pm 18$ & $530\pm 29$ & $178\pm 17$ \\
Fake Lepton & $10\pm 3$ & $48\pm 10$ & $7\pm 2$ & $25\pm 5$ \\ \hline
SUBTRACTED & $65\pm 15\pm 20$ & $35\pm 18\pm 16$ & 
$523\pm 29\pm 38$ & $153\pm 17\pm 13$ \\ \hline   
$\epsilon(\pi^0\ell^+\nu$ MC)\% & $1.01\pm 0.05\pm 0.03$ & 
$0.66\pm 0.05\pm 0.01$ 
& - & - \\
$\epsilon(\overline{K^0}\ell^+\nu$ MC)\% & $0.020\pm 0.004$ & 
 $0.007\pm 0.005$ & $0.54\pm 0.01\pm 0.02$ & $0.17\pm 0.02\pm 0.01$ \\ 
$\epsilon(\overline{K^{*0}}\ell^+\nu$ MC)\% & $<0.001$ & $<0.001$ & 
 $<0.001$ & $<0.001$ \\ \hline 
YIELD ($\times 10^3$)& $6.44\pm 1.49\pm 2.00$ & 
$5.30\pm 2.73\pm 2.84$ & 
$96.85\pm 5.37\pm 7.95$ & $90.00\pm 10.00\pm 13.66$  
\end{tabular}
}
\end{table}

\vfill\eject

%\vfill
%\eject

\newpage
\makebox{}

\newpage
\makebox{}

%\vfill\eject

\end{document}